\documentclass[12pt,letterpaper]{article}
\usepackage[top=1in,left=1in,bottom= 1in,right=1.0in, headheight=14pt]{geometry} 
\usepackage{hyperref}

\usepackage{enumitem}	
\usepackage[scaled=0.92]{helvet}	% set Helvetica as the sans-serif font
\usepackage{siunitx}
\usepackage{array}
\usepackage{longtable}
\usepackage{setspace}

\usepackage{amsmath}		
\usepackage{float}	
\usepackage{graphicx}	
\usepackage{caption} % For captions (optional but recommended)

\newcounter{appendixfigure}

\usepackage{titlesec}
\titleformat{\section}
  {\normalfont\large\bfseries} % Large and bold font
  {\thesection}{1em}{\titlecase}
  
\titleformat{\subsection}
  {\normalfont\normalsize\itshape} % Normal size, italicized, normal font
  {\thesubsection}{1em}{\titlecase} % Capitalize the first letter
  
\titleformat{\subsubsection}
  {\normalfont\normalsize\scshape} % Small capitals, normal size, normal font
  {\thesubsubsection}{1em}{}
  
 \newcommand{\titlecase}[1]{\expandafter\MakeUppercase#1}
\usepackage{array}
\usepackage{titlesec}
\usepackage{cleveref}

\usepackage{cite}

\usepackage{nameref,hyperref}

\usepackage{fancyhdr}
\usepackage{setspace}
\usepackage{booktabs}
\usepackage{longtable}% Set the beginning of a LaTeX document
\usepackage{circuitikz}
\begin{document}

\title{Material- and geometry-independent multishell cloaking device}         % Enter your title between curly braces
\author{ Pattabhiraju C. Mundru\textsuperscript{1}, Venkatesh Pappakrishnan\textsuperscript{1},
and
Dentcho A. Genov\textsuperscript{1}\\
\textsuperscript{\small\bf 1} {\small Department of Physics, Louisiana Tech University, 201 Mayfield Ave, Ruston, LA 71272, USA}    }   % Enter your name between curly braces
\date{June 16, 2012}
\maketitle

\begin{abstract}

In this paper we proposed a multi-shell generic cloaking system. A transparency condition independent of the object’s optical and geometrical properties is proposed in the quasi-static regime of operation. The suppression of dipolar scattering is demonstrated in both cylindrically and spherically symmetric systems. A realistic tunable-low loss shell design is proposed based on composite metal-dielectric shell. The effects due to dissipation and dispersion on the overall scattering cross-section are thoroughly evaluated. It is shown that a strong reduction of scattering by a factor of up to \(10^3\) can be achieved across the entire optical spectrum. Full wave numerical simulations for complex shape particle are performed validating the analytical theory. The proposed design does not require optical magnetism and is generic in the sense that it is independent of the object’s material and geometrical properties.

\end{abstract}
\section{Introduction}
\label{sec:intro}
% TODO: write your article here.
Recently, cloaking or invisibility has attracted significant attention from the scientific community. Several approaches have been proposed to cloak macroscopic and microscopic objects, among which transformation optics (TO)~\cite{Leonhardt2006,Pendry2006,Schurig2006,Cummer2006,Cai2007,Lai2009,Chen2010,McCall2011} and scattering-cancellation techniques based on the elimination of dipolar scattering~\cite{Alu2005,Kerker1975,ZhouHu2006} are the most widely studied. Both approaches rely on engineering one or more material shells surrounding an object in order to render it invisible to external observers.

In 2006, Leonhardt~\cite{Leonhardt2006} employed optical conformal mapping to design an isotropic cloaking device with a spatially varying refractive index surrounding an object, thereby achieving invisibility within the ray-optics approximation. In the same year, Pendry \emph{et al.}~\cite{Pendry2006} introduced the transformation optics framework for designing electromagnetic metamaterial (EMM) cloaks capable of guiding electromagnetic waves around an object while preserving the incident wavefronts. Using the TO formalism, Lai \emph{et al.}~\cite{Lai2009} proposed a complementary-medium cloak capable of concealing objects located outside the cloaking region. McCall \emph{et al.}~\cite{McCall2011} extended the transformation concept to both space and time and proposed a spacetime cloak (STC) capable of concealing events rather than physical objects. Unlike conventional spatial cloaks, which redirect electromagnetic waves around a finite spatial region, an STC manipulates the local velocity of light before and after the event to be hidden, thereby creating a temporal gap in which the event remains undetected. The realization of such devices requires sophisticated temporally modulated metamaterials.

Electromagnetic metamaterials are artificial media engineered to exhibit electromagnetic properties that are difficult or impossible to obtain in naturally occurring materials. Their unusual properties have enabled numerous breakthroughs, including negative-index media~\cite{Valentine2008}, super-resolution imaging~\cite{Fang2005}, cloaking devices~\cite{Schurig2006}, and optical illusion systems that allow one object to appear as another to an external observer~\cite{Lai2009Illusion}. Furthermore, metamaterials have enabled unprecedented control over the propagation of electromagnetic waves,  allowing functionalities that are virtually unattainable using conventional materials~\cite{Milton2006, Genov2009}.

Electromagnetic invisibility through the elimination of dipolar scattering was first investigated by Kerker \emph{et al.}~\cite{Kerker1975} for subwavelength ellipsoidal particles. More recently, Alù and Engheta~\cite{Alu2005} extended this concept to spherical and cylindrical geometries and demonstrated a substantial reduction in the total scattering cross section through the use of dielectric shells with appropriately chosen geometrical and electromagnetic parameters. Zhou \emph{et al.}~\cite{ZhouHu2006} further developed this idea using the concept of neutral inclusions to derive generalized transparency conditions in the quasistatic limit. However, these approaches impose explicit geometrical and material constraints on both the cloaking shells and the object to be concealed. Consequently, the resulting cloaks remain strongly dependent on the optical and geometrical properties of the object, thereby limiting their practical applicability.

In this paper, we propose a multi-shell cloaking strategy capable of concealing an object regardless of its shape or material properties within the quasistatic regime. Transparency conditions that are independent of the object's optical and geometrical characteristics are derived for both cylindrically and spherically symmetric systems. As a practical realization of the proposed cloak, we introduce a tunable, low-loss, near-zero-index shell design based on metal--dielectric composite materials. Our results demonstrate that the proposed structure can achieve cloaking over a broad optical spectral range while significantly reducing the scattering cross section. In addition, full-wave simulations performed for two-dimensional cylindrically symmetric systems confirm the object-independent nature of the design and show excellent agreement with the analytical theory developed herein.

The remainder of this paper is organized as follows. Section~II derives the transparency conditions for a multi-shell cloaking system that is independent of the optical and geometrical properties of the concealed object. Section~III investigates the performance of realistic cloaks composed of dispersive materials, including bulk metals and metal--dielectric composites, and compares their behavior with that of the ideal lossless case. Section~IV presents time-domain and finite-difference frequency-domain (FDFD) analyses of the proposed cloaking system. Finally, conclusions are drawn in Section~V.

\section{Theoretical Analysis}

The geometry of the problem is depicted in Fig.~1. An object of arbitrary shape and permittivity $\varepsilon_0$ is placed inside a cylindrical or spherical domain of radius $r_0$ (core), surrounded by a system of $l$ shells with radii $r_1,r_2,r_3,\ldots,r_l$ $(r_0<r_1<r_2<\cdots<r_l)$ and permittivities $(\varepsilon_0,\varepsilon_1,\varepsilon_2,\ldots,\varepsilon_l)$, respectively. The cloak is embedded in a medium with permittivity $\varepsilon_e$ and illuminated by a uniform electric field $E_0$ polarized along the $+x$-axis (or equivalently by a transverse magnetic (TM) wave). In the quasistatic limit, the electric potential inside and outside the cloak can be written as

\begin{equation}
\varphi_{2D}
=
E_0
\sum_{n=1}^{\infty}
\left(
A_n^{2D} r^n
+
S_n^{2D} r^{-n}
\right)
\cos(n\phi),
\end{equation}

\begin{equation}
\varphi_{3D}
=
E_0
\sum_{n=1}^{\infty}
\left(
A_n^{3D} r^n
+
S_n^{3D} r^{-(n+1)}
\right)
P_n(\cos\theta),
\end{equation}
where $A_n^{d}$ and $S_n^{d}$ are amplitude coefficients, $P_n$ denotes the Legendre polynomials, and $d$ represents the dimensionality of the system. By applying the continuity conditions for the tangential electric field and the normal electric displacement at the interfaces $r_0,r_1,r_2,\ldots,r_l$, the dipolar terms ($n=1$), which are responsible for the far-field scattering in the embedding medium, can be written as
\begin{equation}
S_1^{d}
=
r_l^{d}
\,
\frac{\varepsilon_{\mathrm{eff}}^{\,l}-\varepsilon_e}
{\varepsilon_{\mathrm{eff}}^{\,l}+(d-1)\varepsilon_e},
\end{equation}

where

\begin{equation}
\varepsilon_{\mathrm{eff}}^{\,l}
=
\varepsilon_l
+
\frac{
d\,\varepsilon_l\,p_l
\left(
\varepsilon_{\mathrm{eff}}^{\,l-1}-\varepsilon_l
\right)
}
{
d\,\varepsilon_l
+
(1-p_l)
\left(
\varepsilon_{\mathrm{eff}}^{\,l-1}-\varepsilon_l
\right)
},
\end{equation}

is the effective permittivity of the $l$-shell system, and
\(
p_l
=
\left({r_{l-1}}/{r_l}\right)^d
\) is the surface (cylindrical case) or volume (spherical case) ratio associated with the $l$th shell.
It is important to note that, together with the natural condition
\(
\varepsilon_{\mathrm{eff}}^{\,0}=\varepsilon_0,
\) Eqs.~(3) and (4) provide a straightforward recurrence relation for estimating the scattering coefficient of multilayer dielectric particles in the quasistatic approximation without explicitly solving the complete boundary-value problem (see Appendix~A for details). This formulation also provides an intuitive physical interpretation of the scattering process: the multilayer structure behaves as an equivalent spherical or cylindrical particle with effective permittivity $\varepsilon_{\mathrm{eff}}^{\,l}$ immersed in a host medium of permittivity $\varepsilon_e$.

Alù \emph{et al.}~\cite{Alu2005} and Zhou and Hu~\cite{ZhouHu2006} demonstrated that, in the quasistatic limit, complete suppression of dipolar scattering can be achieved through an appropriate choice of the shell radii. Following their approach, in the limit
\(S_1^{d}\rightarrow 0\)
\(
\left(
\varepsilon_{\mathrm{eff}}^{\,l}\neq\varepsilon_e
\right)\), 
we obtain a generalized transparency condition for an $l$-shell cloaking system that depends explicitly on the object's permittivity and size:
\begin{equation}
p_l
=
\left(
\frac{\varepsilon_l-\varepsilon_{\mathrm{eff}}^{\,l-1}}
{\varepsilon_l-\varepsilon_{\mathrm{eff}}^{\,l}}
\right)
\left(
\frac{\varepsilon_{\mathrm{eff}}^{\,l-1}
+(d-1)\varepsilon_l}
{\varepsilon_e+(d-1)\varepsilon_l}
\right),
\qquad l\geq 1.
\end{equation}
The condition given by Eq.~(5) is consistent with the transparency relations reported in Refs.~\cite{Alu2005,ZhouHu2006} for single-shell and two-shell geometries. In particular, for $l=1$, corresponding to a single-shell spherically symmetric cloak, Eq.~(5) reduces to the transparency condition reported by Alù \emph{et al.}~\cite{Alu2005}.
\begin{equation}
p_1
=
\left(\frac{r_0}{r_1}\right)^d
=
\frac{(\varepsilon_1-\varepsilon_e)\left(2\varepsilon_1+\varepsilon_0\right)}
     {(\varepsilon_1-\varepsilon_0)\left(2\varepsilon_1+\varepsilon_e\right)},
\end{equation}
where $\varepsilon_1$ is the shell permittivity. As evident from Eq.~(6), this design does not require large refractive indices or optical magnetism, in contrast to transformation-optics (TO) cloaks~\cite{Pendry2006}. However, practical realizations of such cloaking systems suffer from a significant limitation: any change in the object's properties requires a complete redesign of the cloak. Specifically, the transparency condition depends explicitly on the object's permittivity $\varepsilon_0$ and radius $r_0$. Furthermore, the approach is applicable only to objects possessing spherical or cylindrical symmetry.

Alternatively, we propose a different transparency condition capable of achieving complete suppression of dipolar scattering for a cloak composed of $l\geq 2$ shells. Inspection of Eqs.~(3) and (4) shows that this condition is satisfied when
\(
\varepsilon_{\mathrm{eff}}^{\,l}=\varepsilon_e,
\)
which can be achieved if the two outermost cloaking shells satisfy
\begin{equation}
\varepsilon_{l-1}=0,
\qquad
\varepsilon_l
=
\varepsilon_e
\frac{1+p_l/(d-1)}
     {1-p_l}.
\end{equation}

Provided that a near-zero-index material can be realized, the permittivity of the outermost shell depends only on the radii of the $l$th and $(l-1)$th shells. This result is in sharp contrast with the conventional transparency condition given by Eqs.~(5) and (6), which depends explicitly on the optical properties of the concealed object. Consequently, in the quasistatic limit, a cloaking system designed according to Eq.~(7) has the potential to cloak objects with arbitrary optical properties.

Furthermore, as demonstrated in Section~IV, the condition $\varepsilon_{l-1}\rightarrow 0$ permits cloaking of objects with arbitrary shapes, provided that they are fully enclosed within shells of order lower than $l-1$. Interestingly, a close similarity exists between the cloak designs proposed here and those based on conventional transformation optics. In ideal transformation-optics cloaks, perfect invisibility is achieved when the constitutive parameters become singular at the inner boundary of the cloak~\cite{Leonhardt2006,Pendry2006}. In the present approach, the near-zero-index shell plays an analogous role by effectively decoupling the concealed object from the incident electromagnetic field.

In transformation optics, perfect cloaking can be achieved provided that the permittivity and/or permeability of the anisotropic shell approaches zero at the boundary between the cloak and the concealed object~\cite{LeonhardtPhilbin2006}. Zero-index materials correspond to a regime in which the local electromagnetic field experiences little or no phase accumulation while propagating through the medium. In the context of cloaking, this behavior implies an effectively infinite local wavelength $(\lambda \rightarrow \infty)$ and, consequently, an effective object size approaching zero.

This observation provides a physical explanation for why both transformation-optics cloaks and the transparency condition proposed in Eq.~(7) operate independently of the geometrical and material properties of the concealed object. An object with an effective size approaching zero interacts only weakly with the incident electromagnetic field and therefore produces negligible scattering. For the cloaking strategy proposed here, the simplest implementation corresponds to a two-shell design.

\section{Cloak Design and Tunability}

\subsection{Metallic Shell Design}

We consider two-shell cylindrical and spherical cloaking systems (see Figs.~1(b) and 1(c)) embedded in air $(\varepsilon_e = 1)$. To satisfy the transparency condition given by Eq.~(7), the inner shell is chosen to be metallic. The permittivity of the metal is described by the Drude model,
\(\varepsilon_m(\omega)
=
\varepsilon_m'(\omega)
+
i\varepsilon_m''(\omega)
=
\varepsilon_b
-
\left({\omega_p^2}/
{\omega(\omega+i\omega_\tau)}\right)
\),  where $\omega_p$ is the plasma frequency, $\omega_\tau$ is the relaxation frequency, and $\varepsilon_b$ accounts for contributions arising from interband electronic transitions.
Clearly, at the modified plasma frequency
\(
\omega_p'
=
{\omega_p}/{\sqrt{\varepsilon_b}},
\)
the real part of the metal permittivity vanishes,
\(
\varepsilon_m'(\omega_p')=0
\), \(
(\omega_\tau/\omega_p' \ll 1),
\)
and the metallic shell can therefore be used to realize the proposed cloak.
To characterize the cloaking performance, we define the relative scattering length (RSL) for the cylindrical case $(d=2)$ and the relative scattering cross section (RSCS) for the spherical case $(d=3)$ as
\begin{equation}
\sigma_R^{d}
=
\frac{\sigma_{\mathrm{cloak}}^{d}}
     {\sigma_{\mathrm{object}}^{d}}
=
\left|
\frac{S_{1,\mathrm{cloak}}^{d}}
     {S_{1,\mathrm{object}}^{d}}
\right|^{2},
\end{equation}
where $\sigma_{\mathrm{cloak}}^{d}$ and $\sigma_{\mathrm{object}}^{d}$ denote the scattering length (for $d=2$) or scattering cross section (for $d=3$) of the cloak and the uncloaked object, respectively.
Substituting the transparency condition given by Eq.~(7) into Eq.~(3), and using
\(\varepsilon_2
=
\varepsilon_m(\omega_p')
=
i\varepsilon_m''(\omega_p')
\) at the modified plasma frequency $\omega_p'$ (where $\varepsilon_m''$ denotes the imaginary part of the metal permittivity), and retaining terms up to second order in the small parameter
\(\varepsilon_m''
=
\varepsilon_b
{\omega_\tau}/{\omega_p'}\),
\(
\left({\omega_\tau}/{\omega_p'} \ll 1\right)\),
we obtain
\begin{equation}
\sigma_R^{d}(\omega_p')
=
\left(
\frac{
d\,\varepsilon_m''(\omega_p')
}
{p_1}
\right)^2
\left[
\frac{
(\varepsilon_0+d-1)\left[1+(d-1)p_1\right]
}
{
(\varepsilon_0-1)(1-p_1)(p_2+d-1)^2
}
\right]^2.
\end{equation}
The presence of material dissipation clearly affects the cloak performance, with
\(\sigma_R^{d}
\propto
\left({\omega_\tau}/{\omega_p}\right)^2
\), increasing quadratically with the relaxation rate $\omega_\tau$.
Furthermore, inspection of Eq.~(9) shows that a geometrical optimization can be achieved, with the scattering reaching a minimum when
\(p_1={1}/{\left(1+\sqrt{d}\right)}
\). Using Eq.~(7), the minimum attainable RSL/RSCS can then be written as a function of the outer-shell permittivity,
\begin{equation}
\sigma_{R,\min}^{d}(\omega_p')
=
\varepsilon_b^{\,3}
\left(
\frac{\omega_\tau}{\omega_p}
\right)^2
\left[
\frac{
(\varepsilon_0+d-1)
\left(d-1+\varepsilon_2^{-1}\right)^2
}
{
d(\varepsilon_0-1)(1-\sqrt{d})^2
}
\right]^2.
\end{equation}

To demonstrate the performance of the proposed cloak, we consider separate cylindrical and spherical designs. Owing to its relatively low dissipative losses, silver is chosen as the metallic constituent, with material parameters
\(
\omega_p = 9.1~\mathrm{eV}
\), \(\omega_\tau = 0.02~\mathrm{eV}
\), and \(
\varepsilon_b = 5.22
\). 
The cloak performance, evaluated as a function of the outer-shell permittivity $\varepsilon_2$, is presented in Fig.~2. For comparison, we also include the corresponding RSL/RSCS of the  ideal lossless system and for a cloak incorporating a silver inner shell are also shown in Fig.~2. For the ideal system, the RSL/RSCS approaches zero, indicating perfect suppression of scattering at the outer-shell permittivities $\varepsilon_2=5$ and $\varepsilon_2=10$ for the cylindrical and spherical geometries, respectively. These values are in excellent agreement with those predicted by the transparency condition given by Eq.~(7).
A substantial reduction in scattering is also observed for the metallic-shell cloak at the predicted values of the outer-shell permittivity $\varepsilon_2$ (see Fig.~2, dashed curves). For comparison, the minimum attainable RSL and RSCS predicted by Eq.~(10) are also included in Fig.~2. These curves provide a useful estimate of the maximum cloaking performance achievable with the proposed design. At the optimal outer-shell permittivity, the exact numerical results are in excellent agreement with the analytical predictions. Furthermore, for strongly scattering objects $(\varepsilon_0\rightarrow\infty)$, the minimum scattering asymptotically approaches the limit
\(\sigma_{R,\mathrm{RSCS}}^{\min}
\to
\varepsilon_b^{\,3}
\left({\omega_\tau}/{\omega_p}\right)^2
\left(1+{1}/{\sqrt{d}}\right)^4 \ll 1
\). Finally, it should be noted that cloaks based on metallic shells inherently possess a relatively narrow operational bandwidth. The operating frequency range
\(
\omega \in (\omega_p' - \Delta\omega,\;\omega_p' + \Delta\omega)
\)
is approximately determined by the condition
\(
\left|
\varepsilon_m''(\omega_p'+\Delta\omega)
\right|
=
\left|
\varepsilon_m''(\omega_p'-\Delta\omega)
\right|,
\)
which yields
\(
\Delta\omega={\omega_\tau}/{2}.
\)

\subsection{Composite Media: Optical Tunability}

From the discussion in the previous subsection, it is evident that although metallic shells can provide a substantial reduction in scattering at the modified plasma frequency, their response cannot be tuned over a broad spectral range. Such tunability, however, can be achieved through nanocomposite materials consisting of metallic inclusions with permittivity $\varepsilon_m$ embedded in a dielectric host of permittivity $\varepsilon_h$~\cite{Garcia2002,Genov2003,MaxwellGarnett1904}.

If the inclusions are  randomly oriented ellipsoidal inclusions with a small volume fraction $f$, the effective permittivity of the composite medium is given by the Maxwell--Garnett formula~\cite{BergmanStroud1992,Genov2003,MaxwellGarnett1904,Polder1946},
\begin{equation}
\varepsilon_{\mathrm{eff}}
=
\varepsilon_h
+
\frac{f}{3}
\sum_{j=x,y,z}
\frac{
\varepsilon_h(\varepsilon_m-\varepsilon_h)
}
{
\varepsilon_h+\eta_j(\varepsilon_m-\varepsilon_h)
},
\end{equation}
where $\eta_j$ are the depolarization factors of the ellipsoidal inclusions satisfying the relation
\(
\sum_{j=x,y,z}\eta_j = 1.
\)

For prolate spheroids (needle-shaped inclusions), i.e., ellipsoids with semi-axes $a>b=c$, the depolarization factor along the major axis is given by
\begin{equation}
\eta_x=\eta
=
\frac{1-e^2}{e^2}
\left[
\frac{1}{2e}
\ln\!\left(\frac{1+e}{1-e}\right)
-1
\right],
\end{equation}
where
\(
e^2=1-\left({b}/{a}\right)^2
\)
is the eccentricity~\cite{Genov2003}. The remaining depolarization factors are identical and given by
\(\eta_y=\eta_z={\left(1-\eta\right)}/{2}\). 

To reduce dissipative losses and simplify the design, we consider only the low-frequency resonance of the ellipsoidal inclusions, determined by the condition
\(\varepsilon_m'(\omega)
=
-\varepsilon_h
{\left(1-\eta\right)}/{\eta}
\). The effective permittivity of the composite depends on both the material properties and the geometry of the inclusions, thereby providing considerable flexibility in satisfying the transparency condition given by Eq.~(7). Since Eq.~(11) is valid for small inclusion concentrations, typically $f<5\%$, we investigate the effects of varying the depolarization factor $\eta$ (i.e., the shape of the metallic inclusions), the host permittivity $\varepsilon_h$, or both. For small metallic losses,
\(
{\varepsilon_m''}/{\varepsilon_h}\ll 1,
\)
the depolarization factor satisfying the transparency condition
\(
\varepsilon_{\mathrm{eff}}'(\omega_{op})=0
\)
can be approximated by

\begin{equation}
\eta
=
\frac{1}
{1-\varepsilon_m'(\omega_{op})/\varepsilon_h}
-\frac{f}{3}.
\end{equation}
where $\omega_{\mathrm{op}}$ denotes the operating frequency, and the depolarization factor must vary from $\eta=0$ (needle-like inclusions) to $\eta=1/3$ (spherical inclusions). Correspondingly, for a given depolarization factor, the operating frequency is obtained as
\(\omega_{\mathrm{op}}
=
{\omega_p}/\left(
{\sqrt{\varepsilon_b-\varepsilon_h+{3\varepsilon_h}/{(3\eta+f)}}}\right).
\)It should be noted that Eq.~(13) is valid only when
\(
f
>
6\varepsilon_h\varepsilon_m''
\left[
(\varepsilon_m')^2
+
(\varepsilon_h-\varepsilon_m')^2
\right]^{-1}
\),
otherwise the transparency condition cannot be satisfied, i.e.,
\(
\varepsilon_{\mathrm{eff}}' > 0.
\)

The operating frequency range of the composite cloak is shown in Fig.~3(a). For air as the host medium, tuning the depolarization factor of the ellipsoidal inclusions yields an operational frequency range of
\(
\hbar\omega_{\mathrm{op}}
\in
(1.14,\,3.89)\ \mathrm{eV}.
\)
For the same inclusion volume fraction but with glass as the host medium, the operating frequency exhibits a red shift and falls within the range
\(
\hbar\omega_{\mathrm{op}}
\in
(0.83,\,3.74)\ \mathrm{eV}
\).
Thus, by varying the host material and the aspect ratio of the ellipsoidal inclusions (equivalently, the depolarization factor), the operating frequency can be tuned over a substantial portion of the visible and near-infrared spectrum.
However, very small depolarization factors correspond to highly elongated spheroids, which may become impractical for cloak implementation. In addition, the physical dimensions of the composite structure must remain much smaller than the incident wavelength in order for the effective-medium approximation to remain valid. Consequently, we impose the constraint
\(
{1}/{3} \geq \eta \geq {1}/{10}
\), which corresponds approximately to spheroidal aspect ratios in the range \(1 \leq {a}/{b} < 3\). This restriction is sufficiently weak to permit effective cloaking across the entire optical spectral range considered in this work.

To estimate the RSL/RSCS of the composite cloak, we employ Eq.~(9) with the substitution
\(
\varepsilon_m''
\rightarrow
\varepsilon_{\mathrm{eff}}''(\omega_{\mathrm{op}}),
\) where the effective permittivity of the composite is obtained from Eqs.~(11) and (13) and is given by
\begin{equation}
\varepsilon_1
=
i\,\varepsilon_{\mathrm{eff}}''(\omega_{\mathrm{op}})
=
\frac{
3\varepsilon_h^{\,2}
\,i\varepsilon_m''
}
{
f\left(\varepsilon_h-\varepsilon_m'\right)^2
}.
\end{equation}
For the geometrically optimized design, corresponding to\(p_1={1}/{\left(1+\sqrt{d}\right)}
\), low-frequency operation and strongly scattering objects $(\varepsilon_0\rightarrow\infty)$, the minimum RSCS/RSL asymptotically approaches
\begin{equation}
\sigma_{R,\min}^{d}(\omega_{\mathrm{op}})
=
\left(
\frac{
3\omega_\tau\omega_p\,\varepsilon_h^{\,2}
}
{
f\,\omega_p^{2}
}
\right)^2
\left(
\frac{1+\sqrt{d}}
{\sqrt{d}}
\right)^4.
\end{equation}

Figures~3(b) and 3(c) show the RSCS/RSL as a function of the outer-shell permittivity $\varepsilon_2$ for two-shell cylindrical and spherical cloaks with inner shells composed of different metal--dielectric composites. For comparison, results for the ideal lossless case are also included, together with composite hosts consisting of air (dotted curves) and glass (dashed curves).

For all cases, the operating frequency is fixed at
\(
\hbar\omega_{\mathrm{op}}=1.14~\mathrm{eV}
\).
A substantial reduction in scattering is observed for all composite-based designs, with the air-host composite exhibiting superior performance, as predicted by Eq.~(15). Furthermore, the composite cloak provides approximately a $40\%$ reduction in scattering compared with designs employing bulk-metal inner shells (see Fig.~2).

As in the previous cases, the minimum RSCS and RSL occur at values of the outer-shell permittivity $\varepsilon_2$ satisfying the transparency condition given by Eq.~(7). The figures also include the geometrically optimized predictions obtained from Eqs.~(9) and (14), which are in excellent agreement with the exact numerical calculations.

Finally, to provide a more complete characterization of the cloak performance, Figs.~4(a) and 4(b) present the RSCS/RSL as functions of the outer-shell permittivity $\varepsilon_2$ and the operating frequency $\omega_{\mathrm{op}}$, respectively. A substantial reduction in both the RSCS and RSL is observed throughout the visible and near-infrared spectral regions. As predicted by Eq.~(15), the scattering increases with increasing 
frequency to the point at which the effect of the cloak on the scattering cross section is no longer beneficial, i.e.,
\(
\sigma_{R,\min}^{d}>1\), 
\(\hbar\omega_{\mathrm{op}}>3.2~\mathrm{eV}\).
It should be noted that a further reduction in scattering may be achieved by increasing the volume fraction of the metallic spheroids, provided that the validity condition of Eq.~(11) remains satisfied.
\section{Full-Wave Analysis of a Generic Cylindrical Cloak}
The transparency condition given by Eq.~(7) is derived within the quasistatic approximation and is therefore valid only for objects whose physical dimensions are much smaller than the wavelength of the incident radiation. As the object size becomes comparable to the wavelength, the quasistatic approximation breaks down and the transparency condition is expected to fail. To investigate this transition and establish the size limitations of the proposed design, we perform a full-wave analysis of a two-shell cylindrical cloak operating at optical and near-infrared frequencies.

We consider the scattering of a transverse magnetic (TM) plane wave by an infinitely long two-shell cylindrical cloak, as illustrated in Fig.~1(b). The incident and scattered magnetic fields outside the cloak can be expressed in the standard form
\begin{equation}
H_z
=
H_z^{i}
+
H_z^{s}
=
H_0
\sum_{n=-\infty}^{\infty}
i^n
\left[
J_n(k_e r)
+
S_n H_n^{(1)}(k_e r)
\right]
e^{in\phi},
\end{equation}
where $J_n$ and $H_n^{(1)}$ are the Bessel and Hankel functions of the first kind, respectively,
\(k_e=\left({\omega}/{c}\right)\sqrt{\varepsilon_e}
\)
is the wave number in the host medium, and $S_n$ are the scattering coefficients obtained by applying the appropriate boundary conditions at the shell interfaces (see Appendix~A).

The scattering cross length is then given by the standard multipole expansion~\cite{BohrenHuffman1998,KerkerMatijevic1961,Mie1908}
\begin{equation}
\sigma
=
\frac{4}{k_e}
\sum_{n=-\infty}^{\infty}
|S_n|^2.
\end{equation}

In the calculations, the geometrical parameters of the cloak are chosen according to the optimal design,
\(
p_1={1}/{\left(1+\sqrt{d}\right)},
\) \(
p_2=0.67,
\)
and the shell permittivities are selected to satisfy the transparency conditions given by Eqs.~(7) and (14).

Figure~5 illustrates the RSL of the composite cloak for cylindrical dielectric and metallic particles serving as the concealed object. As expected, for systems with small overall dimensions [see Figs.~5(a) and 5(c)], a substantial reduction in scattering across the entire optical spectrum is achieved for
\(
\varepsilon_2=\left({1+p_2}\right)/\left({1-p_2}\right),
\)
thereby reproducing the quasistatic result. Compared with a dielectric particle, a low RSL over a broader frequency range is observed for a metallic object. This behavior arises from the strong enhancement of the metallic particle scattering at the surface-plasmon frequency
\(\hbar\omega_{\mathrm{sp}}
=
{\omega_p}/{\left(\sqrt{\varepsilon_b+\varepsilon_e}\right)}
=
3.71~\mathrm{eV}.
\)
However, as the system size increases [see Figs.~5(b) and 5(d)], the transparency condition given by Eq.~(7) is no longer sufficient to suppress the scattering process. This behavior is expected because the contribution of higher-order multipoles, for which
\(
k_e r_2 \geq 1,
\) 
increases with the physical size of the structure (see Appendix~A).
For higher-order cylindrical multipoles, the scattering coefficients are approximately
\begin{equation}
S_n
=
\frac{i\pi}
{\Gamma(n)\Gamma(n+1)}
\left(\frac{k_e r_2}{2}\right)^{2n}
\frac{
\varepsilon_2(1-p_2^{\,n})-\varepsilon_e(1+p_2^{\,n})
}{
\varepsilon_2(1-p_2^{\,n})+\varepsilon_e(1+p_2^{\,n})
}.
\end{equation}

It is straightforward to show that the quadrupolar scattering contribution of the cloak becomes comparable to, and eventually exceeds, the dipolar scattering contribution of the object when
\(
k_e r_2
\geq
\sqrt{
8p_1
\left(
1+p_2+p_2^2
\right)
},
\)
assuming $\varepsilon_0 \gg 1$.
Overall, for particle diameters larger than approximately $400~\mathrm{nm}$, a substantial reduction in scattering cannot generally be expected throughout the optical spectral range.

Finally, we address one of the most important features of the proposed cloak, namely its independence from the optical and geometrical properties of the concealed object. The condition
\(
\varepsilon_{l-1}=0
\)
implies
\(\varepsilon_{\mathrm{eff}}^{\,l-1}=0,
\)
as follows directly from Eq.~(4), regardless of the effective permittivity $\varepsilon_{\mathrm{eff}}^{\,l-2}$ of the underlying shell--object substructure. Consequently, the cloaking response becomes independent of the properties of the concealed object. This feature allows the proposed design to cloak objects of virtually arbitrary shape and composition, provided that they are encapsulated within at least two shells $(l\geq2)$.

To verify this property, full-wave simulations were performed using the finite difference frequency domain (FDFD) solver implemented in \textsc{COMSOL Multiphysics}. A metallic rounded star-shaped object was placed inside the cloak. The shell permittivities were chosen as
\(
\varepsilon_2=5
\), \(
\varepsilon_1=0,
\)
with a radius ratio
\(
p_1=0.67,
\)
corresponding to
\(
r_2=106~\mathrm{nm},
\) and \(
r_1=87~\mathrm{nm}.
\)
The system was illuminated by a TM-polarized cylindrical wave generated by a point source located $130~\mathrm{nm}$ from the center of the object.

The resulting magnetic-field distribution is shown in Fig.~6. In the presence of the cloak, the cylindrical wave generated by the source smoothly propagates around the cloaked region, indicating a substantial reduction in scattering [Fig.~6(a)]. The phase fronts remain nearly undisturbed as they exit the cloak, and no significant shadow formation is observed.

For comparison, Fig.~6(b) shows the magnetic-field distribution after removing the cloak. In this case, the incident wave is strongly scattered by the object, and the phase fronts become severely distorted after interacting with it. Pronounced shadowing and resonant field enhancements within the object are clearly visible.

The striking difference between the two field distributions demonstrates that the cloaking strategy based on the transparency condition given by Eq.~(7) can substantially reduce scattering from objects with diverse optical and geometrical properties. These results provide numerical evidence supporting the object-independent nature of the proposed cloaking design.

\section{Summary}

In this work we propose a generic cloaking system
based on zero-permittivity composite materials. The proposed
analytical model and full wave calculations show that a
dramatic suppression of dipolar scattering can be achieved
for an arbitrary object enclosed within a multishell cloaking
system. A reduction of scattering across the entire optical
spectrum for dielectric objects using realistic shell materials
is demonstrated. This study provides a direction for achieving
optical invisibility without the use of metamaterials and also
underlines the role of zero-index materials in the general
phenomenon of optical transparency
\section*{Acknowledgements}

The authors would like to thank Dr.~S.~G.~Moiseev, Dr.~N.~Simicevic, Dr.~A.~Sihvola, Dr.~D.~Robbins, K.~Inturi, and S.~Animilli for many useful discussions.

This work was supported by the Louisiana Board of Regents and the National Science Foundation under contracts LEQSF (2007--12)-ENH-PKSFI-PRS-01, LEQSF (2011--14)-RD-A-18, and NSF (2010)-PFUND-202.

\bibliography{mundrupappagenov.bib}

@article{Cai2007,
  author = {Cai, W. and Chettiar, U. K. and Kildishev, A. V. and Shalaev, V. M.},
  title = {Optical Cloaking with Metamaterials},
  journal = {Nature Photonics},
  volume = {1},
  pages = {224--227},
  year = {2007}
}

@article{Cummer2006,
  author = {Cummer, S. A. and Popa, B. I. and Schurig, D. and Smith, D. R. and Pendry, J. B.},
  title = {Full-wave simulations of electromagnetic cloaking structures},
  journal = {Physical Review E},
  volume = {74},
  pages = {036621},
  year = {2006}
}

@article{Leonhardt2006,
  author = {Leonhardt, U.},
  title = {Optical Conformal Mapping},
  journal = {Science},
  volume = {312},
  pages = {1777--1780},
  year = {2006}
}

@article{Pendry2006,
  author = {Pendry, J. B. and Schurig, D. and Smith, D. R.},
  title = {Controlling Electromagnetic Fields},
  journal = {Science},
  volume = {312},
  pages = {1780--1782},
  year = {2006}
}

@article{LeonhardtPhilbin2006,
  author = {Leonhardt, U. and Philbin, T.},
  title = {General Relativity in Electrical Engineering},
  journal = {New Journal of Physics},
  volume = {8},
  pages = {247},
  year = {2006}
}

@article{Lai2009,
  author = {Lai, Y. and Chen, H. and Zhang, Z.-Q. and Chan, C. T.},
  title = {Complementary Media Invisibility Cloak that Cloaks Objects at a Distance Outside the Cloaking Shell},
  journal = {Physical Review Letters},
  volume = {102},
  pages = {093901},
  year = {2009}
}

@article{Chen2010,
  author = {Chen, H. and Chan, C. T. and Sheng, P.},
  title = {Transformation Optics and Metamaterials},
  journal = {Nature Materials},
  volume = {9},
  pages = {387--396},
  year = {2010}
}

@article{McCall2011,
  author = {McCall, M. W. and Favaro, A. and Kinsler, P. and Boardman, A.},
  title = {A Spacetime Cloak, or a History Editor},
  journal = {Journal of Optics},
  volume = {13},
  pages = {024003},
  year = {2011}
}

@article{Schurig2006,
  author = {Schurig, D. and Mock, J. J. and Justice, B. J. and Cummer, S. A. and Pendry, J. B. and Starr, A. F. and Smith, D. R.},
  title = {Metamaterial Electromagnetic Cloak at Microwave Frequencies},
  journal = {Science},
  volume = {314},
  pages = {977--980},
  year = {2006}
}

@article{Valentine2008,
  author = {Valentine, J. and Zhang, S. and Zentgraf, T. and Ulin-Avila, E. and Genov, D. A. and Bartal, G. and Zhang, X.},
  title = {Three-dimensional Optical Metamaterial with a Negative Refractive Index},
  journal = {Nature},
  volume = {455},
  pages = {376--379},
  year = {2008}
}

@article{Fang2005,
  author = {Fang, N. and Lee, H. and Sun, C. and Zhang, X.},
  title = {Sub-Diffraction-Limited Optical Imaging with a Silver Superlens},
  journal = {Science},
  volume = {308},
  pages = {534--537},
  year = {2005}
}

@article{Lai2009Illusion,
  author = {Lai, Y. and Ng, J. and Chen, H.-Y. and Han, D. and Xiao, J. and Zhang, Z.-Q. and Chan, C. T.},
  title = {Illusion Optics: The Optical Transformation of an Object into Another Object},
  journal = {Physical Review Letters},
  volume = {102},
  pages = {253902},
  year = {2009}
}

@article{Genov2009,
  author = {Genov, D. A. and Zhang, S. and Zhang, X.},
  title = {Mimicking Celestial Mechanics in Metamaterials},
  journal = {Nature Physics},
  volume = {5},
  pages = {687--692},
  year = {2009}
}

@article{Alu2005,
  author = {Alu, A. and Engheta, N.},
  title = {Achieving Transparency with Plasmonic and Metamaterial Coatings},
  journal = {Physical Review E},
  volume = {72},
  pages = {016623},
  year = {2005}
}

@article{Kerker1975,
  author = {Kerker, M.},
  title = {Invisible Bodies},
  journal = {Journal of the Optical Society of America},
  volume = {65},
  pages = {376},
  year = {1975}
}

@article{ZhouHu2006,
  author = {Zhou, X. and Hu, G.},
  title = {Design for Electromagnetic Wave Transparency with Metamaterials},
  journal = {Physical Review E},
  volume = {74},
  pages = {026607},
  year = {2006}
}

@article{Milton2006,
  author = {Milton, G. W. and Nicorovici, N.-A. P.},
  title = {On the Cloaking Effects Associated with Anomalous Localized Resonance},
  journal = {Proceedings of the Royal Society A},
  volume = {462},
  pages = {3027--3059},
  year = {2006}
}

@article{Garcia2002,
  author = {Garcia, N. and Ponizovskaya, E. V. and Xiao, J. Q.},
  title = {Zero Refractive Index in Nanostructured Materials},
  journal = {Applied Physics Letters},
  volume = {80},
  pages = {1120--1122},
  year = {2002}
}

@incollection{BergmanStroud1992,
  author = {Bergman, D. J. and Stroud, D.},
  title = {Physical Properties of Macroscopically Inhomogeneous Media},
  booktitle = {Solid State Physics},
  volume = {46},
  pages = {147--269},
  publisher = {Academic Press},
  year = {1992}
}

@article{Genov2003,
  author = {Genov, D. A. and Sarychev, A. K. and Shalaev, V. M.},
  title = {Metal-Dielectric Composite Materials},
  journal = {Journal of Nonlinear Optical Physics and Materials},
  volume = {12},
  pages = {419--428},
  year = {2003}
}

@article{MaxwellGarnett1904,
  author = {Maxwell Garnett, J. C.},
  title = {Colours in Metal Glasses and Metallic Films},
  journal = {Philosophical Transactions of the Royal Society of London A},
  volume = {203},
  pages = {385--420},
  year = {1904}
}

@article{Polder1946,
  author = {Polder, D. and Van Santen, J. H.},
  title = {The Effective Permeability of Mixtures of Solids},
  journal = {Physica},
  volume = {12},
  pages = {257--271},
  year = {1946}
}

@book{BohrenHuffman1998,
  author = {Bohren, C. F. and Huffman, D. R.},
  title = {Absorption and Scattering of Light by Small Particles},
  publisher = {Wiley},
  address = {New York},
  year = {1998}
}

@article{KerkerMatijevic1961,
  author = {Kerker, M. and Matijevic, E.},
  title = {Scattering of Light by Colloidal Particles},
  journal = {Journal of the Optical Society of America},
  volume = {51},
  pages = {506--508},
  year = {1961}
}

@article{Mie1908,
  author = {Mie, G.},
  title = {Beitr{\"a}ge zur Optik Tr{\"u}ber Medien},
  journal = {Annalen der Physik},
  volume = {330},
  pages = {377--445},
  year = {1908}
}

\bibliographystyle{abbrv}

\newpage

\begin{figure}[ht]         % Starts the floating environment with placement specifiers
    \centering               % Centers the image horizontally
    \includegraphics[width=0.7\textwidth]{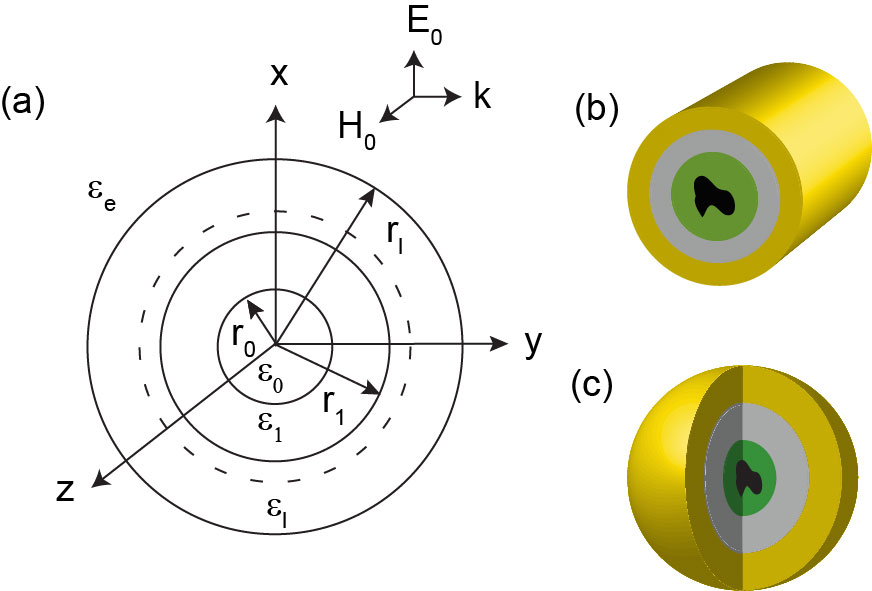} % Loads and sizes the graphic
    \caption{ (a) Generic multi-shell cloaking system with shell radii $r_i$ and permittivity $\varepsilon_i$, a two shell (b) cylindrically symmetric cloaking system and (c) spherically symmetric cloaking system. }   % Adds an automated number and caption
    \label{fig:my_image}     % Creates a hidden marker for text referencing
\end{figure}

\newpage

\begin{figure}[H]         % Starts the floating environment with placement specifiers
    \centering               % Centers the image horizontally
    \includegraphics[width=0.5\textwidth]{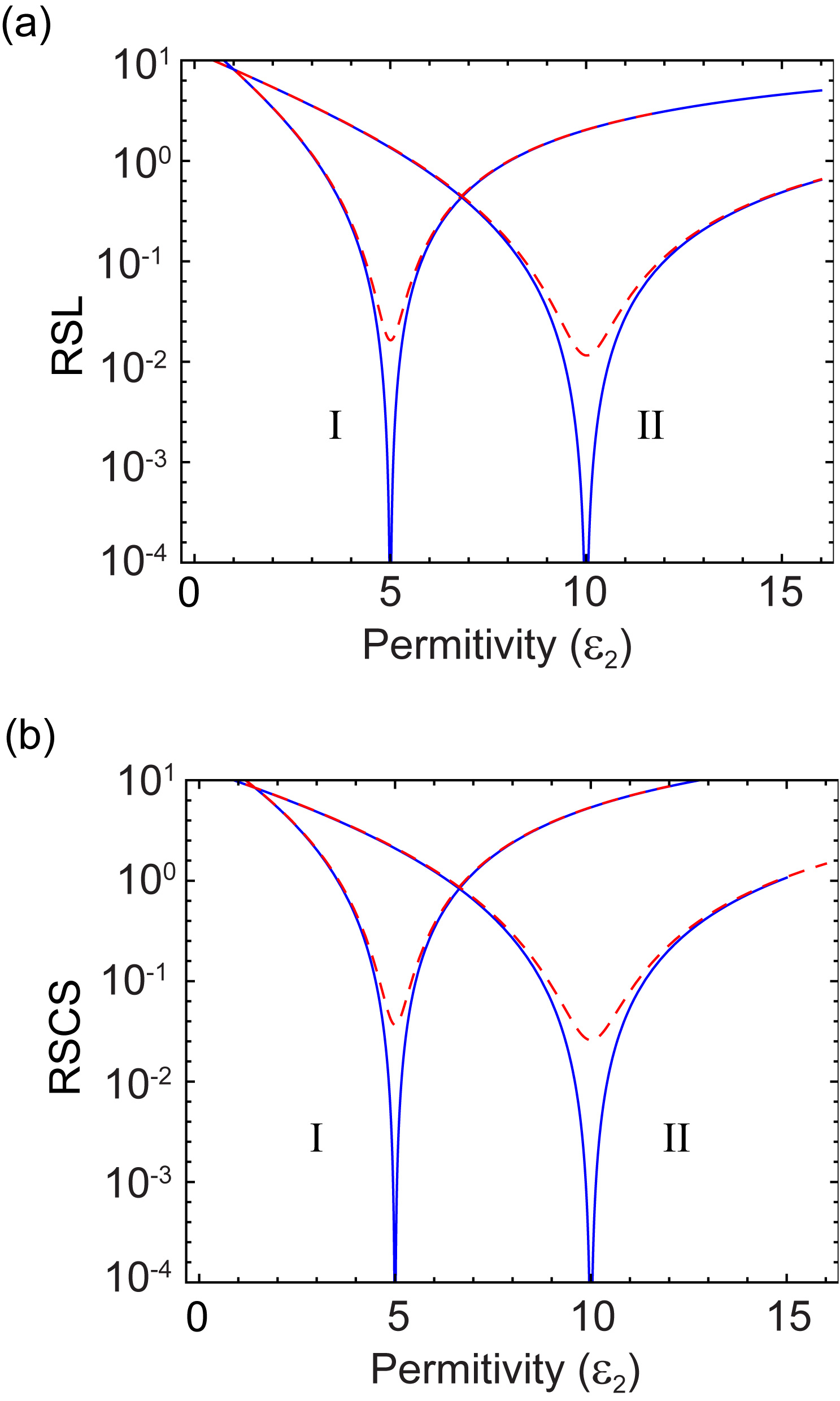} % Loads and sizes the graphic
   \caption{
Relative scattering length (RSL) and relative scattering cross section (RSCS) versus outer-shell permittivity for a two-shell cloak with a bulk-silver inner shell. (a) Cylindrical geometry: (I) $p_2=0.67$ and (II) $p_2=0.82$. (b) Spherical geometry: (I) $p_2=0.73$ and (II) $p_2=0.86$. The object permittivity is $\varepsilon_0=12$ and the optimal shell ratio $p_1=1/(1+\sqrt{d})$ is used throughout. Black dotted lines indicate the limiting values predicted by Eq.~(10).
}% Adds an automated number and caption
    \label{fig:fig2}     % Creates a hidden marker for text referencing
\end{figure}
\newpage

\begin{figure}[ht]         % Starts the floating environment with placement specifiers
    \centering               % Centers the image horizontally
    \includegraphics[width=1.0\textwidth]{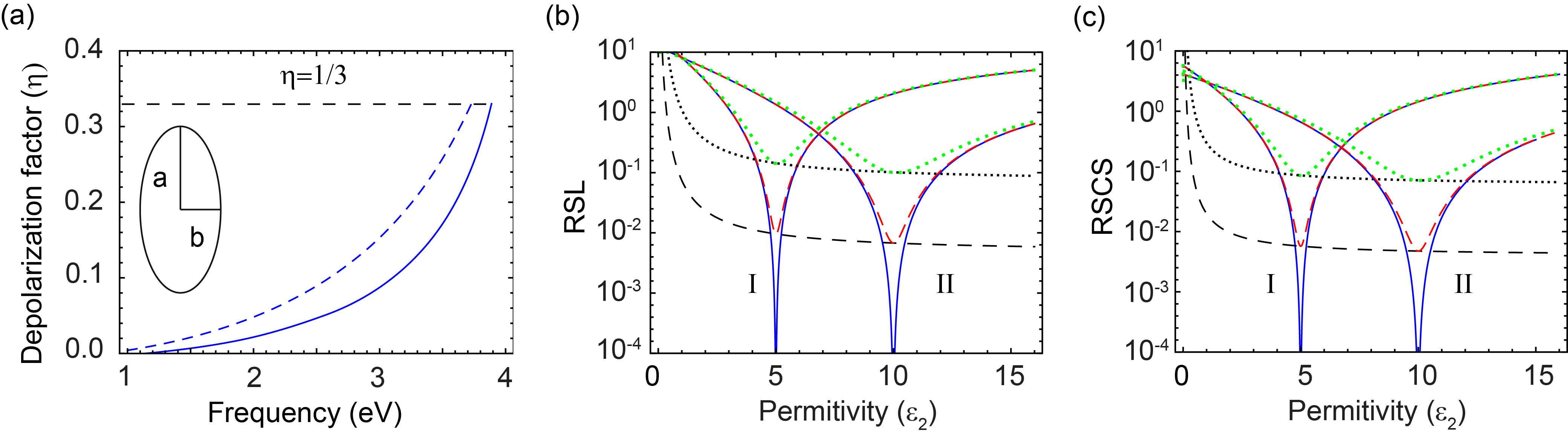} % Loads and sizes the graphic
    \caption{
(a) Depolarization factor versus frequency for composite hosts with $\varepsilon_h=1$ (solid) and $\varepsilon_h=2$ (dashed), assuming $f=0.05$. Relative scattering length (RSL) and relative scattering cross section (RSCS) versus outer-shell permittivity for a two-shell cloak with a composite inner shell: (b) cylindrical geometry and (c) spherical geometry. The composite shell is designed using host materials $\varepsilon_h=1$ (red dashed curve) and $\varepsilon_h=2$ (green dotted curve) at $\hbar\omega=1.14\,\mathrm{eV}$. Cylindrical designs: (I) $p_2=0.67$, (II) $p_2=0.82$. Spherical designs: (I) $p_2=0.73$, (II) $p_2=0.86$. The object permittivity is $\varepsilon_0=12$ and the optimal shell ratio $p_1=1/(1+\sqrt{d})$ is used throughout. Solid blue curves denote the ideal lossless case. Horizontal black dashed and dotted lines indicate the limiting values predicted by Eqs.~(9) and (14).
}  % Adds an automated number and caption
    \label{fig:fig3}     % Creates a hidden marker for text referencing
\end{figure}

\newpage

\begin{figure}[ht]         % Starts the floating environment with placement specifiers
    \centering               % Centers the image horizontally
    \includegraphics[width=0.5\textwidth]{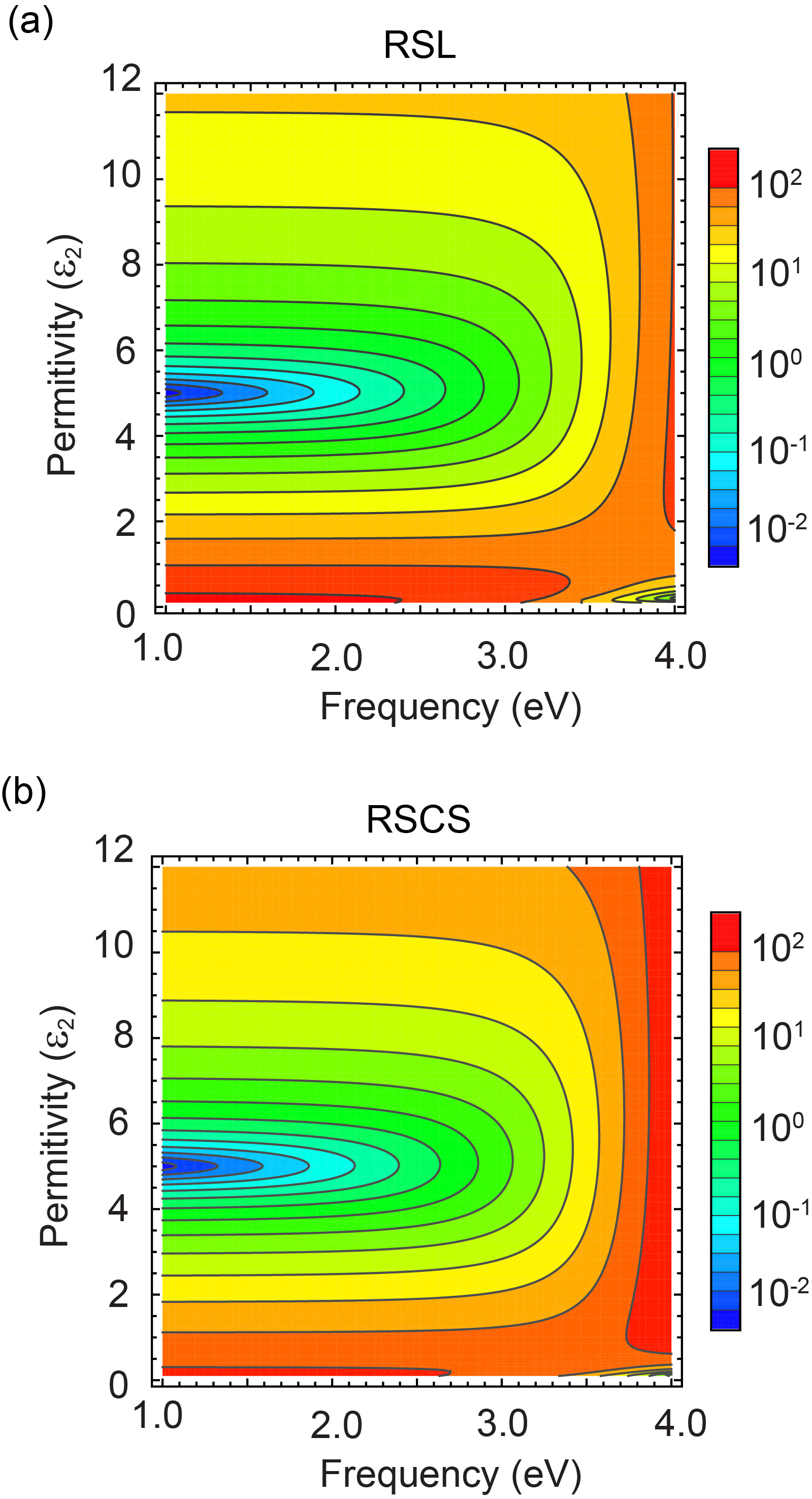} % Loads and sizes the graphic
    \caption{
Relative scattering length (RSL) and relative scattering cross section (RSCS) as functions of outer-shell permittivity $\varepsilon_2$ and incident-light frequency for a dielectric object with $\varepsilon_0=12$. The composite inner shell is characterized by $\varepsilon_h=1$ and spheroidal inclusion volume fraction $f=0.05$. (a) Cylindrical cloak with $(p_1,p_2)=(0.41,0.67)$. (b) Spherical cloak with $(p_1,p_2)=(0.37,0.73)$.
} % Adds an automated number and caption
    \label{fig:fig4}     % Creates a hidden marker for text referencing
\end{figure}

\newpage

\begin{figure}[ht]         % Starts the floating environment with placement specifiers
    \centering               % Centers the image horizontally
    \includegraphics[width=1.0\textwidth]{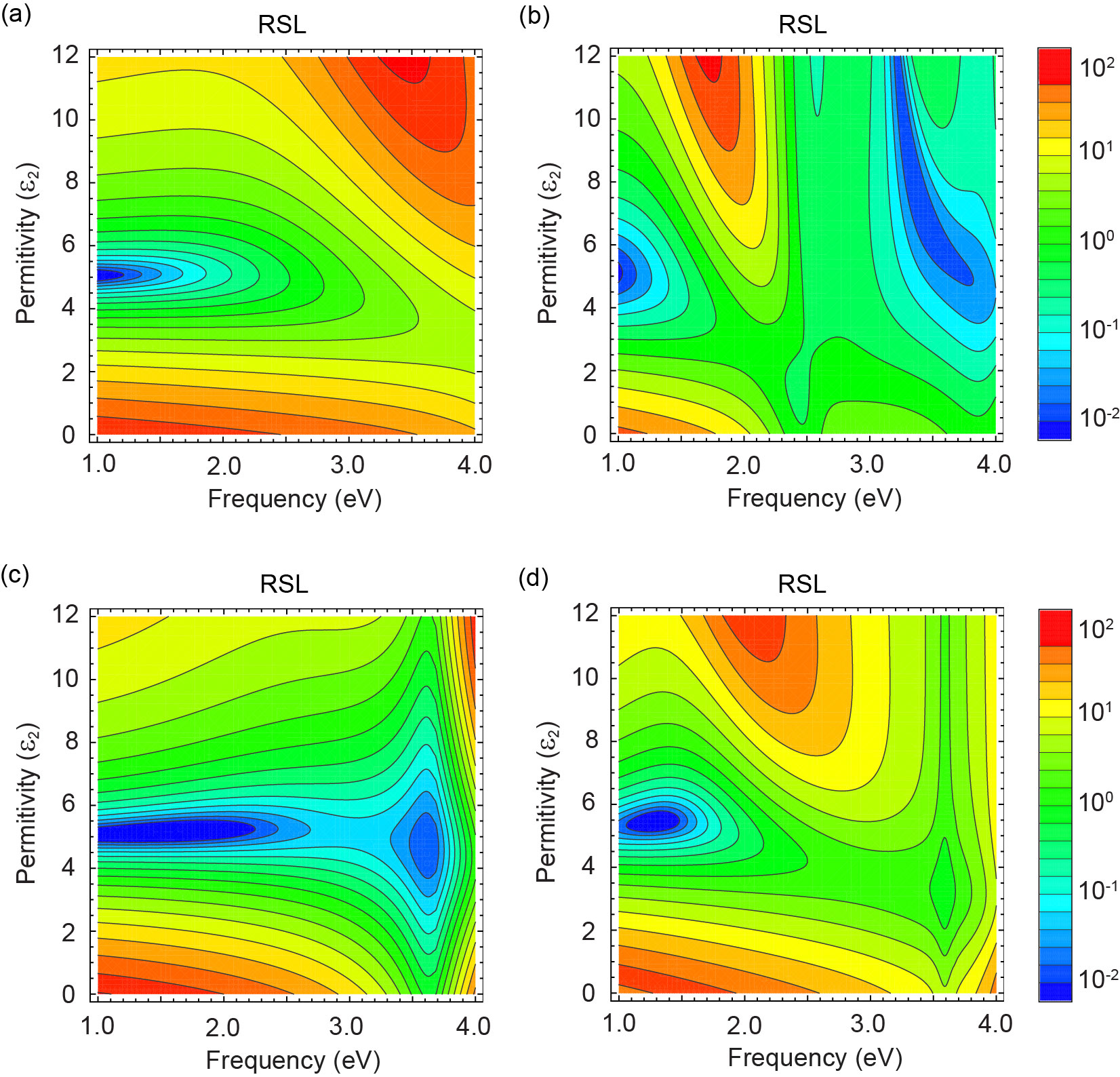} % Loads and sizes the graphic
    \caption{
Full-wave calculations of the relative scattering length (RSL) versus outer-shell permittivity $\varepsilon_2$ and incident-light frequency. Results are shown for a dielectric particle ($\varepsilon_0=12$) in panels (a) and (b), and for a metallic particle in panels (c) and (d). The cloak radius is $r_2=50\,\mathrm{nm}$ for (a) and (c), and $r_2=100\,\mathrm{nm}$ for (b) and (d).
} % Adds an automated number and caption
    \label{fig:fig5}     % Creates a hidden marker for text referencing
\end{figure}

\newpage

\begin{figure}[ht]         % Starts the floating environment with placement specifiers
    \centering               % Centers the image horizontally
    \includegraphics[width=0.5\textwidth]{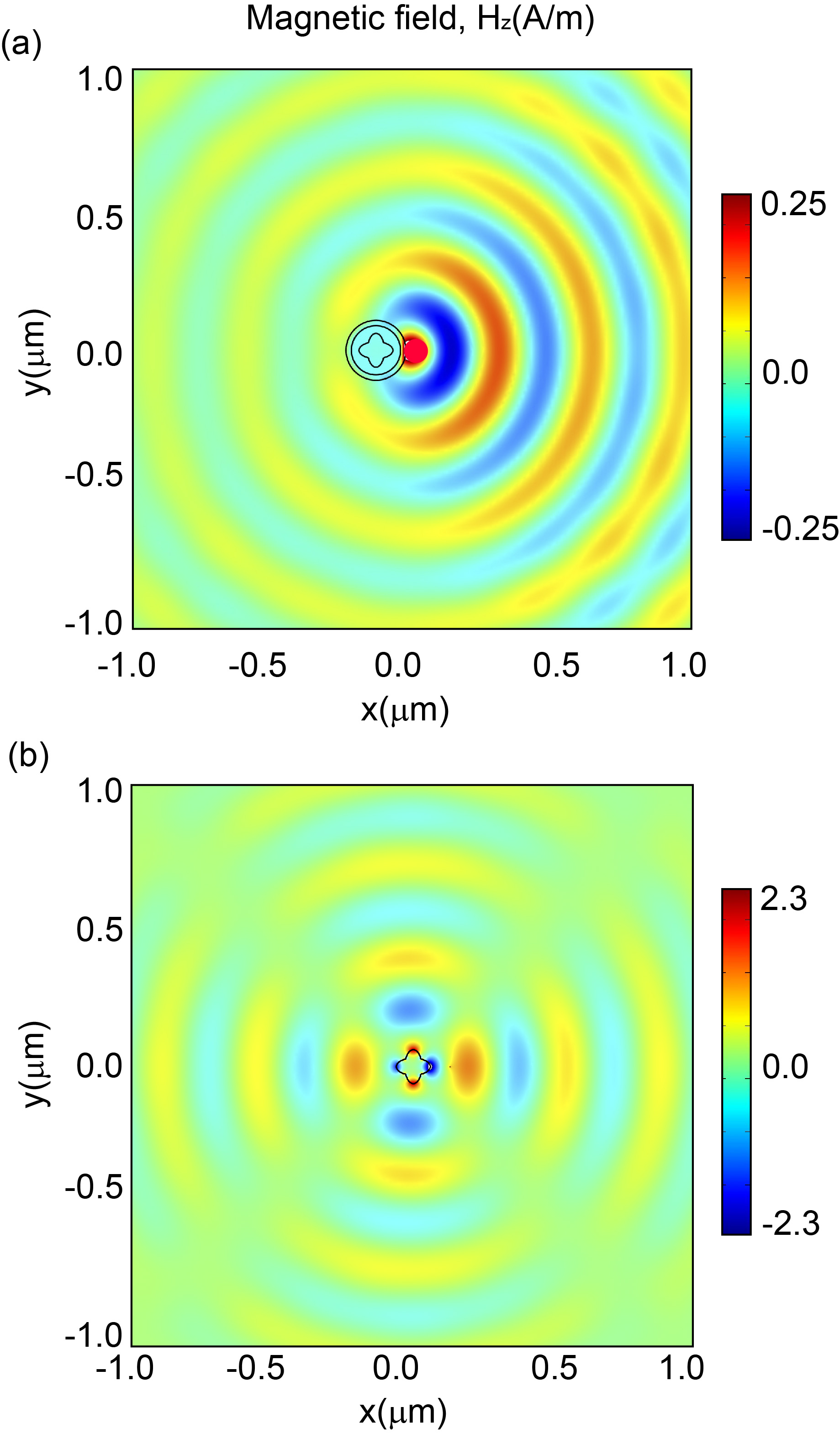} % Loads and sizes the graphic
 \caption{
Full-wave simulations of light scattering from a point source near a star-shaped metallic object: (a) with the cloak and (b) without the cloak. The incident TM-polarized wave has photon energy $\hbar\omega = 3.8\,\mathrm{eV}$.
}% Adds an automated number and caption
    \label{fig:fig6}     % Creates a hidden marker for text referencing
\end{figure}

\end{document}